# One Monad to Prove Them All

Sandra Dylus[a], Jan Christiansen[b], and Finn Teegen[a]

a   University of Kiel, Germany
b   Flensburg University of Applied Sciences, Germany

**Abstract**   One Monad to Prove Them All is a modern fairy tale about curiosity and perseverance, two important properties of a successful PhD student. We follow the PhD student Mona on her adventure of proving properties about Haskell programs in the proof assistant Coq.

On the one hand, as a PhD student in computer science Mona observes an increasing demand for correct software products. In particular, because of the large amount of existing software, verifying existing software products becomes more important. Verifying programs in the functional programming language Haskell is no exception. On the other hand, Mona is delighted to see that communities in the area of theorem proving are becoming popular. Thus, Mona sets out to learn more about the interactive theorem prover Coq and verifying Haskell programs in Coq.

To prove properties about a Haskell function in Coq, Mona has to translate the function into Coq code. As Coq programs have to be total and Haskell programs are often not, Mona has to model partiality explicitly in Coq. In her quest for a solution Mona finds an ancient manuscript that explains how properties about Haskell functions can be proven in the proof assistant Agda by translating Haskell programs into monadic Agda programs. By instantiating the monadic program with a concrete monad instance the proof can be performed in either a total or a partial setting. Mona discovers that the proposed transformation does not work in Coq due to a restriction in the termination checker. In fact the transformation does not work in Agda anymore as well, as the termination checker in Agda has been improved.

We follow Mona on an educational journey through the land of functional programming where she learns about concepts like free monads and containers as well as basics and restrictions of proof assistants like Coq. These concepts are well-known individually, but their interplay gives rise to a solution for Mona's problem based on the originally proposed monadic tranformation that has not been presented before. When Mona starts to test her approach by proving a statement about simple Haskell functions, she realizes that her approach has an additional advantage over the original idea in Agda. Mona's final solution not only works for a specific monad instance but even allows her to prove monad-generic properties. Instead of proving properties over and over again for specific monad instances she is able to prove properties that hold for all monads representable by a container-based instance of the free monad. In order to strengthen her confidence in the practicability of her approach, Mona evaluates her approach in a case study that compares two implementations for queues. In order to share the results with other functional programmers the fairy tale is available as a literate Coq file.

If you are a citizen of the land of functional programming or are at least familiar with its customs, had a journey that involved reasoning about functional programs of your own, or are just a curious soul looking for the next story about monads and proofs, then this tale is for you.

## The Art, Science, and Engineering of Programming

**Prologue**  This is a literate Coq file Dylus [11], that is, it can be compiled with Coq[1] as it is.

## 1  Preamble

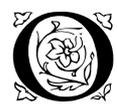NCE upon a time there was a computer scientist called Mona D. living in the beautiful land of functional programming. The land of functional programming was divided into several kingdoms. Mona lived in the kingdom of Haskell, a land where all citizens used the pure lazy functional programming language Haskell.

Mona was pursuing her PhD when one day, her adviser told her about the magical kingdom of Coq. In the magical kingdom of Coq, all citizens used the dependently typed programming language Coq. With Coq, the citizens were not only able to write functional programs but also to formally prove properties about these programs.

Mona was fascinated by the idea of proving properties about programs. While her colleagues in the department of software engineering were still writing tests and hoped to catch all bugs that way, she could once and for all prove that her programs were correct. Mona's adviser suggested to her to visit an allied department where she could stay for a semester. With her good intentions at heart, our hero set out to the kingdom of Coq to learn more about the Coq programming language.

Soon after Mona got to the kingdom of Coq she realized that proving statements about programs came at a price in Coq. In Coq, all functions have to be provably total. Most of the citizens in the kingdom of Haskell, however, did not care about totality as they were used to working with partial functions. Therefore, Mona could not simply use the Haskell programs she had one-to-one in Coq.

Our hero went to the largest library in the kingdom of Coq and consulted many manuscripts[2] written by sages from the entire land of functional programming. The first manuscript she found was written by Breitner et al. [5]. In this manuscript, the authors proved statements about Haskell programs in Coq. They transformed Haskell programs into Coq programs by keeping as much of the structure of the Haskell programs as possible. In order to model partial functions they added an opaque constant *default* : *A* for every non-empty type *A*. By means of this constant they defined a function *error* : *String* → *A* that was used to model the corresponding Haskell function. Because *default* is opaque, the concrete value of *default* cannot be used in any proofs.

While Mona really liked the manuscript by Breitner, Spector-Zabusky, Li, Rizkallah, Wiegley, and Weirich, they could not reason about partial values and, thus, about the strictness of an implementation. Furthermore, she found manuscripts that emphasized the importance of considering partial values. For example, in a manuscript by Johann et al. [15] she read about the *destroy/unfoldr* rule as presented by Svenningsson [23].

---

[1] The file was tested with Coq version 8.7 and 8.8.
[2] In Mona's days the term manuscript was used as a synonym for all kinds of scientific references.

This rule fuses a list consumer with a list producer. The *destroy/unfoldr* rule does not hold if partial values are considered and a strict evaluation primitive like seq is available. Moreover, partial functions are often necessary in the implementation of data structures, whose invariants cannot be encoded in Haskell's type system. For example, a function might be undefined for an empty list but an invariant states that the argument will never be an empty list.

Mona also read a manuscript by Danielsson et al. [10] about relating the total interpretation of a function with an interpretation of the same function that models partiality. In essence, when the function is only applied to total arguments, the two interpretations behave the same. While Mona saw the importance of this connection, the approach was still not able to argue about partial functions. For example, it was not sufficient that a program transformation held for all total arguments, if the compiler could not enforce that the arguments are always total.

Mona did not give up. There had to be a way to get the best of both worlds. That is, argue about a Haskell function with respect to total values whenever possible and argue about partial values only if necessary. Mona went to the largest library in the whole land of functional programming. And indeed she found a manuscript by Abel et al. [1] that discussed exactly the problem she was trying to solve. Abel, Benke, Bove, Hughes, and Norell lived in the nearby magical kingdom of Agda, where dependently typed programming was also highly regarded by its citizens. Abel, Benke, Bove, Hughes, and Norell presented an approach to translate Haskell programs into monadic Agda programs. By instantiating the monadic program with a concrete monad instance one could choose the model a statement should be proven in. For example, if Mona chose the identity monad, she would work in a total world while the maybe monad allowed her to argue about partiality. Furthermore, she would not have to make any changes to her program but could argue about the same program regardless of the monad she chose.

Abel, Benke, Bove, Hughes, and Norell used monadic expressions to model possibly undefined expressions. As Haskell is non-strict, a data structure may contain an undefined expression that is never demanded. For example, a cons may have an undefined tail. Mona enthusiastically defined the following inductive data type for lists in Coq, which resembled the corresponding definition by Abel, Benke, Bove, Hughes, and Norell in Agda.[3]

*Fail* Inductive *List M A* :=
| *nil* : *List M A*
| *cons* : $M\ A \to M\ (List\ M\ A) \to List\ M\ A$.

In Coq, all arguments of a type or function can be annotated: values as well as types. As a Haskell programmer, Mona was not used to annotating type variables with their types, so she used the code convention that type variables would not be annotated, if they could be inferred — just like she was used to in her Haskell programs. That is, Mona could use $M$ : Type $\to$ Type and $A$ : Type in the definition of *List*, but these annotations were inferred by the system, so she left them out. Moreover, she used the convention that constructors start with a lower-case letter.

---

[3] Coq uses GADT-style data type definitions.

When Mona tried to compile the definition of *List* in a current version of Coq, she was quite disappointed. As indicated by the keyword *Fail*, the Coq compiler accepted the definition *List* only if it caused an error. In Mona's case, the Coq compiler presented the following error message.

> The command has indeed failed with message:
> Non-strictly positive occurrence of "*List*" in "*M A* → *M (List M A)* → *List M A*".

Mona thought that she had made a mistake when translating the Agda definition to Coq. She wrote a letter to a friend in the kingdom of Agda. A few days later she received an unexpected answer: the definition did not work in Agda either. At the time the manuscript by Abel et al. [1] was written, the rules in the kingdom of Agda were less strict, but that was a long time ago. Meanwhile, the king of the kingdom of Agda had realized that the rules needed to be changed. The programs that could be defined using the old rules were not safe! Following the example of the magical kingdom of Coq, non-strictly positive occurrences as mentioned in the error message above were now disallowed in the kingdom of Agda as well.

In this story, Mona D. went on a journey to find a model for Haskell programs in Coq that obeyed the king's rules. She learned about the laws of the kingdom of Coq from the perspective of a programmer from the kingdom of Haskell. Mona would also meet fellows with an interest in more theoretical aspects of functional programs and approaches for generic programming in Haskell. On the way to achieving her goal, there were many obstacles to overcome. Led by her curiosity and her perseverance, reading a lot of manuscripts and meeting with her fellows would ultimately give Mona the inspiration how to tackle her problem.

## 2 The Problem

Mona spent the next days searching for manuscripts about possible adaptations of the approach by Abel et al. [1] that worked in a current version of Coq. Nothing came up, but Mona did not want to give up. Thus, she used her most powerful weapon in defeating scientific obstacles: her perseverance. She tried multiple simplified versions of data type definitions similar to *List* and checked for each whether the error still occurred. The simplest data type definition she found that caused the same error looked as follows.

*Fail* Inductive *NonStrictlyPos* := *con* : (*NonStrictlyPos* → *nat*) → *NonStrictlyPos*.

When Mona compiled this definition, she got the following error message.

> The command has indeed failed with message:
> Non-strictly positive occurrence of "*NonStrictlyPos*" in "(*NonStrictlyPos* → nat) → *NonStrictlyPos*".

So Mona wondered what it meant that *NonStrictlyPos* occurs in a non-strictly positive way in the type signature of *con* : (*NonStrictlyPos* → *nat*) → *NonStrictlyPos*. Mona remembered hearing about a class of Adam Chlipala explaining the intricacies of Coq very well. She looked up the term "strictly positive" and indeed found the relevant note in Chlipala's manuscript [6].

> We have run afoul of the strict positivity requirement for inductive definitions, which says that the type being defined may not occur to the left of an arrow in the type of a constructor argument.

An inductive type like *NonStrictlyPos* obeys the strict positivity requirement, if its recursive occurrences are strictly positive in the types of all arguments of its constructors. Mona realized that the important part of the definition was that she needed to check the types of all arguments of a constructor and not the type of the constructor itself. Here, the inductive type has only one constructor *con* and this constructor has only one argument of type *NonStrictlyPos* → *nat*. A type $\tau$ occurs strictly positively in a type $\tau_1 \to ... \to \tau_n \to \tau$ if and only if $\tau$ does not occur in any of the types $\tau_i$ with $i$ ranging from 1 to $n$. Mona took another look at the malicious definition and understood. The type *NonStrictlyPos* occurs non-strictly positively in one of its argument types, namely in *NonStrictlyPos* → *nat*. The recursive occurrence of the type *NonStrictlyPos* is "left of an arrow". In other words *NonStrictlyPos* occurs as an argument type of *NonStrictlyPos* → *nat*, which makes *NonStrictlyPos* a non-strictly positive type. In contrast, the following definition is not problematic.

Inductive *StrictlyPos* := *con* : *StrictlyPos* → (*nat* → *StrictlyPos*) → *StrictlyPos*.

The first argument of *con* of type *StrictlyPos* is not problematic as it is trivially strictly positive. In the second argument of *con*, *StrictlyPos* does not occur to the left of an arrow in the type *nat* → *StrictlyPos*. That is, as *StrictlyPos* occurs strictly positively in all arguments of the constructor *con*, it fulfills the strict positivity requirement.

Mona still wondered why the king disallowed a definition like *NonStrictlyPos* in Coq. At lunch the next day, she talked about her insights with some PhD students. One of them had read about restrictions of dependently typed languages and explained to her that while the definition itself did not cause any trouble, it is possible to define functions that cause trouble. Back at Mona's computer, they tried to understand the restriction given the following innocent looking function in Coq assuming that the above definition of *NonStrictlyPos* was accepted.

Definition *applyFun* (*t* : *NonStrictlyPos*) : *nat* :=
  match *t* with
  | *con f* ⇒ *f t*
  end.

The function *applyFun* simply took a value of type *NonStrictlyPos* and applied the function inside the argument of *con* to the value itself. A problematic example usage of this function is the expression *applyFun* (*con applyFun*). Reducing the expression by using the definition of *applyFun* yields *applyFun* (*con applyFun*) again, which indicates that this expression will not terminate.

Mona now remembered something she learned in her first course about functional programming in Haskell, namely the following data type.

*Fail* Inductive *Mu A* := *mu* : (*Mu A* → *A*) → *Mu A*.

**One Monad to Prove Them All**

This data type can be used to implement a fix-point combinator without explicit recursion. Before her colleague left to prepare his next tutoring class, he suggested her to consult a manuscript by McAdam [20] to refresh her memory about the *Mu* data type. If Coq allows a data type like *Mu*, all of a sudden general recursion is available and the corresponding logic becomes inconsistent. Thus, a data type definition that has recursive occurrences to the left of an arrow as in *Mu* and *NonStrictlyPos* is not allowed. When Abel, Benke, Bove, Hughes, and Norell published their work, Agda did not perform any termination checks and, as a consequence, did not check the strict positivity requirement.

While Mona now understood why the data type as defined by Abel, Benke, Bove, Hughes, and Norell is not allowed, it was still not obvious to her why the restriction is violated in the case of *List*. For example, if she explicitly used the maybe monad for *M*, everything worked fine. She thought about all the different monad instances she learned in class. Then she remembered the continuation monad and realized that an arbitrary type constructor *M* might violate the restriction. More concretely, she considered the following definition of the continuation data type.

Definition *Cont R A* := $(A \to R) \to R$.

When she instantiated *M* in the definition of *List* with *Cont R* for some *R*, she got the following type that violated the strict positivity requirement.

*Fail* Inductive *ListC R A* :=
| *nilC* : *ListC R A*
| *consC* : $((A \to R) \to R) \to ((ListC\ R\ A \to R) \to R) \to ListC\ R\ A$.

The definition of *consC* is more delicate than the definitions before. Here, *ListC* occurs non-strictly positive in the second argument of the constructor *consC*, namely (*ListC R A* $\to R) \to R$. Mona's colleague referred her to a manuscript by Coquand et al. [8] that introduced the restriction to forbid non-strictly positive types in the kingdom of Coq. He also brought to her attention that technically, although it was not the case, the type *ListC* could be allowed in Agda. In Agda only types with recursive occurrences in negative positions have to be disallowed. Negative positions are positions left of an odd number of arrows. For example, the definition of *NonStrictlyPos* cannot be allowed because the recursive occurrence in *NonStrictlyPos* $\to$ *nat* is left of one arrow. Just as the definition of *Mu* cannot be allowed in Agda either because the recursive occurrence in *Mu A* $\to A$ is left of one arrow. In contrast the recursive occurrence in (*ListC R A* $\to R) \to R$ is left of two arrows and, therefore, could be allowed in Agda. Coq has to be more restrictive because the sort Prop is impredicative while it is predicative in Agda [9].

In summary, the type constructor *List* defined above allows arbitrary type constructors as instances of *M*. That is, it is not safe to use this definition for all potential instantiations of *M*. The restriction might be violated for a concrete instantiation, like for example *Cont* as used in *ListC*. Since it cannot be guaranteed by definition that this data type declaration is used safely, Coq rejects the declaration.

## 3 The Solution

**M**ONA realized that she could not use a type variable in the definition of the monadic list data type, and, thus, had to use a concrete data type. However, she would still like to model several different monad instances in order to be able to reason about several different possible effects.

**Free Monads**

Mona was a regular reader of the *Haskell Weekly News*, a gazette where citizens of the kingdom of Haskell presented usages and developments of the Haskell programming language. She remembered a lot of stories about monadic abstractions and in particular about the free monad. The free monad turns any functor into a monad and is usually defined as a data type with two constructors, *pure* and *impure*. Mona could use the concrete data type definition of a free monad to represent the monadic part in the problematic list definition. Keen to try out this idea, Mona defined the following data type, where the type variable *F* needs to be a functor to make *Free F* a monad.

*Fail* Inductive *Free F A* :=
| *pure* : $A \to Free\ F\ A$
| *impure* : $F\ (Free\ F\ A) \to Free\ F\ A$.

Again, Coq rejected the definition of this data type. In the constructor *impure* the type variable *F* is applied to *Free F A* and, thus, is violating the aforementioned strict positivity requirement. Mona was disappointed that she could not apply the idea to represent the monadic part of the list definition via the free monad to her problem. As she was still a Haskell programmer with all her heart and the free monad was known to have various applications in functional programming, she continued to study free monads as a distraction from her current setback.

As Mona was mainly looking for a way to represent the identity and the maybe monad, she checked how these monads can be represented by instances of *Free*. She found the corresponding definitions in a manuscript that mentioned free monads, written by Swierstra [24]. The identity monad corresponds to a free monad without the *impure* case, because the identity monad does not have any effect. That is, in order to model the identity monad, Mona needed a functor that had no inhabitants. This way, it was not possible to construct an impure value. Mona defined the following Coq data type that has no inhabitants.

Inductive *Void* := .

Based on this definition Mona defined the following functor to model an instance of the free monad without impure values.

Definition *Zero* (*A* : Type) := *Void*.

Since the type variable *A* did not occur on the right-hand side of this definition, Mona had to actually annotate the sort in order to compile the definition in Coq.

In the case of the maybe monad there is an effect, namely the absence of a value, represented by the constructor *Nothing*. That is, Mona searched for a functor such that the *impure* constructor models the *Nothing* constructor. Therefore, she needed a type constructor that has a single constructor but does not contain any value, such that the recursive occurrence of *Free F A* was not used. She defined a functor with a single value by means of the data type *unit*, which contains a single value called *tt*.

Definition *One* (*A* : Type) := *unit*.

At first Mona was mainly interested in the identity and the maybe monad — because Abel et al. [1] also considered these effects to model Haskell programs. However, when she checked out other definitions, she realized that using the free monad opens the door to model a variety of different monads. For example, the error monad could be modeled by using the following functor.

Inductive *Const* (*B A* : Type) := *const* : *B* → *Const B A*.

Mona had fun trying out various functors and checking what monad corresponded to the resulting instance of the free monad. However, after a while the excitement vanished and Mona was kind of disappointed. It felt as if she had not gained much by using a free monad. Then, Mona suddenly had an insight. By using a free monad she had gained one crucial advantage: she would be able to represent strictly positive monads if she was able to represent strictly positive functors.

Motivated by this insight, she gained new impulse. She got in contact with a group of sages in the area of verification stating her problem concerning Coq's restriction. These experts pointed her to a manuscript by Keuchel et al. [16]. Keuchel and Schrijvers wanted to define the following data type in Coq that can be used to define the fix point of a functor.

*Fail* Inductive *FixF F* := *fixF* : *F* (*FixF F*) → *FixF F*.

The type *FixF* is a generalization of *Mu* as defined in section 2. Thus, the data type *FixF* cannot be defined in Coq because the functor *F* might place its argument *FixF F* in a non-strictly positive position. In order to be able to define this data type the functor *F* has to be restricted. This restriction is based on the notion of containers as introduced by Abott et al. [2]. Mona was quite happy about the feedback because she could use the same approach to define *Free*.

**Containers**

The idea of a container abstracts data types that store values. A prominent example is the list data type. A list can be represented by the length of the list and a function mapping positions within the list, that is, mapping natural numbers between one and the length of the list to the elements of the list. This idea can be generalized to other data types like trees. In this case, instead of a simple natural number, one needs a more complex data type to model all possible positions of a specific shape.

A container is given by a type *Shape* that models all possible shapes and a function *Pos* that takes a shape and yields the type of the possible positions of the given shape. The extension of a container is the concrete data type that is modeled, that is, the extension provides a mapping of valid positions to values [3]. Mona used the following implementation of a container extension in Coq.

Inductive *Ext Shape* (*Pos* : *Shape* → Type) *A* := *ext* : ∀ *s*, (*Pos s* → *A*) → *Ext Shape Pos A*.

For example, in the case of lists, the extension is the mapping of list indices to list elements.

Mona learned a lot about Coq by defining the data type *Ext*. For example, she learned that, in Coq — in contrast to Haskell — when a polymorphic constructor or function was applied, the type that was used as an instance was passed explicitly. For example, the constructor *ext* was polymorphic over the type *A*. In order to apply *ext* Mona would have to pass the concrete instance for the type variable *A*. However, it was possible to persuade Coq to infer these types from the corresponding values. Mona tweaked all constructors and functions in her Coq code so that they behaved like constructors and functions in Haskell. For example, while *ext* would normally take five arguments in Coq — the three types *Shape*, *Pos* and *A*, a shape $s$ : *Shape* and a function of type *Pos s* $\to$ *A*, by tweaking it, it took only the last two arguments as a Haskell programmer would expect.

In order to understand the use of *Ext* Mona modeled the data type *One* as a container following the naming convention used by Swierstra [24]. She came up with the following definition.

Definition $Shape_{One}$ := *unit*.

Definition $Pos_{One}$ (s : $Shape_{One}$) := *Void*.

Definition $Ext_{One}$ A := *Ext* $Shape_{One}$ $Pos_{One}$ A.

The data type *One* was polymorphic over its type argument *A*, which was used as a phantom type and, therefore, not used on the right-hand side of the definition. Furthermore, *One* had just one constructor that did not hold any value. As *One* did not contain any values, the container only had a single shape without any positions. Mona modeled the shape of *One* with the unit type. She used the name *Shape* for the type of shapes, as in $Shape_{One}$, and *Pos* for the corresponding position type. The type of the extension of the container, which consists of the shape type $Shape_{One}$ and the position function $Pos_{One}$ as well as the type of the elements *A*, was named $Ext_{One}$.

The container extension $Ext_{One}$ was isomorphic to *One*. In order to prove this property Mona defined two functions to transform these data types into one another. She called these functions $to_{One}$ and $from_{One}$, respectively.

Definition $to_{One}$ A (e : $Ext_{One}$ A) : *One* A := *tt*.

Definition $from_{One}$ A (z : *One* A) : $Ext_{One}$ A := *ext tt* ($\lambda$ p : $Pos_{One}$ tt $\Rightarrow$ match p with end).

The definition of $to_{One}$ was straightforward as there was only a single value of type *One A*. The definition of $from_{One}$ passed its unit value *tt* to the constructor *ext*. The second argument of *ext* was a function that could never be applied because its argument type has no inhabitants. Therefore, the pattern matching on the position *p* could never be successful. As the type of *p* had no inhabitants, the pattern matching did not have a right-hand side.

At last, Mona proved that $to_{One}$ and $from_{One}$ were actually inverse to each other and, thus, formed an isomorphism between $Ext_{One}$ and *One*.

Lemma $to\_from_{One}$ : $\forall$ A (ox : *One* A), $to_{One}$ ($from_{One}$ ox) = ox.

Lemma $from\_to_{One}$ : $\forall$ A (e : $Ext_{One}$ A), $from_{One}$ ($to_{One}$ e) = e.

In order to define a data type that represented all free monads whose functor was a container, Mona defined a type class named *Container*. This type class was parametrized over a type constructor $F$ : Type $\to$ Type that was isomorphic to the corresponding container extension. The type class provided the type of shapes and the mapping of a shape to the type of positions. Furthermore, the type class provided functions *to* and *from* for transition between the functor and the corresponding container extension. As Mona got more familiar with the advantages of Coq, the type class also provided propositions that *from* and *to* formed a bijection.

Class *Container F* :=
  {
    *Shape* : Type;
    *Pos* : *Shape* $\to$ Type;
    *to* : $\forall$ *A*, *Ext Shape Pos A* $\to$ *F A*;
    *from* : $\forall$ *A*, *F A* $\to$ *Ext Shape Pos A*;
    *to_from* : $\forall$ *A* (*fx* : *F A*), *to* (*from fx*) = *fx*;
    *from_to* : $\forall$ *A* (*e* : *Ext Shape Pos A*), *from* (*to e*) = *e*
  }.

In order to complete the example for *One*, Mona defined the following instance named $C_{One}$[4].

Instance $C_{One}$ : *Container One* :=
  {
    *Shape* := $Shape_{One}$;
    *Pos* := $Pos_{One}$;
    *to* := $to_{One}$;
    *from* := $from_{One}$;
    *to_from* := $to\_from_{One}$;
    *from_to* := $from\_to_{One}$
  }.

**Implementation**

Equipped with these new insights about containers and their usage to define strictly positive types, Mona defined the following data type that implements a free monad whose functor is a container. Here and in the following Mona assumed that $F$ : Type $\to$ Type was a type constructor.

  Inductive *Free* ($C_F$ : *Container F*) *A* :=
  | *pure* : *A* $\to$ *Free $C_F$ A*
  | *impure* : *Ext Shape Pos* (*Free $C_F$ A*) $\to$ *Free $C_F$ A*.

Instead of using an arbitrary functor $F$ the definition of *Free* used a container extension to ensure that the definition was provably valid in Coq, i.e., *Free* only occurred in strictly positive positions. To test her definition, Mona implemented the following abbreviation. The value *Nothing*[5] was the value of the free monad that represented the corresponding value of the maybe monad.

Definition *Nothing A* : *Free $C_{One}$ A* := *impure* (*ext tt* ($\lambda$ *p* : $Pos_{One}$ *tt* $\Rightarrow$ match *p* with end)).

---

[4] Mona could define multiple instances for the same type in Coq, because all instances were explicitly named.

[5] She used the convention that smart constructor names started with an upper case character.

When Mona continued reading about free monads, she read that from every natural transformation from a functor *f* to a monad *m* she can construct a monad homomorphism from *Free f* to *m*. Therefore, for every monad *m* there exists a monad homomorphism from an instance of *Free* to *m* because she can use the identity function as natural transformation. More precisely, the simplest construction Mona can use to model a monad *m* is to represent it as *Free m*.

The homomorphism was often defined by means of a fold function for the *Free* data type. One other important function on *Free* that Mona wanted to define was the function bind, which was associated with monads. She found the following definitions of fold and bind on *Free* in the manuscript by Swierstra [24].

```
foldFree :: Functor f => (a -> b) -> (f b -> b) -> Free f a -> b
foldFree pur imp (Pure x) = pur x
foldFree pur imp (Impure fx) = imp (fmap (foldFree pur imp) fx)

(»=) :: Functor f => Free f a -> (a -> Free f b) -> Free f b
Pure x »= f = f x
Impure fx »= f = Impure (fmap (»= f) fx)
```

Based on these definitions, Mona translated the functions to Coq, naming them *fold_free* and >>=, respectively.

*If you are keen to know how Mona translated these functions to Coq, you and Mona can dive deeper into technical details in section A.1. If you know how to work with recursive higher-order definitions and containers, or do not need any more details, just read ahead.*

The function *fold_free* was only a means to an end. Mona defined the function *induce*, which lifts a function that maps the functor $F$ to the corresponding monad $M$[6] to a homomorphism between the free monad and the corresponding monad[7].

Definition *induce A* ($f : \forall X, F\ X \to M\ X$) (*fx* : *Free* $C_F$ *A*) : *M A* :=
  *fold_free* ($\lambda\ x \Rightarrow ret\ x$) ($\lambda\ x \Rightarrow join\ (f\ (M\ A)\ x)$) *fx*.

To test her current setup, Mona implemented the following functions that map instances of *Free* to the identity and the maybe monad, respectively.

Definition *zero_to_id A* (*zx* : *Zero A*) : *Id A* := match *zx* with end.

Definition *free_to_id A* (*fx* : *Free* $C_{Zero}$ *A*) : *Id A* := *induce zero_to_id fx*.

Definition *one_to_maybe A* (*ox* : *One A*) : *Maybe A* := *nothing*.

Definition *free_to_maybe A* (*fx* : *Free* $C_{One}$ *A*) : *Maybe A* := *induce one_to_maybe fx*.

The functions *zero_to_id* and *one_to_maybe*, respectively, map the functors *Zero* and *One* to the corresponding monads. As there are no values of type *Zero A*, Mona used an empty pattern match to get a value of type *Id A* to define *zero_to_id*.

Mona was quite satisfied with her development so far, using containers she successfully defined a valid type definition for free monads in Coq. However, Mona also realized an important point she had not considered before. Namely, while for every monad *m* there exists a functor *f* with a homomorphism from *Free f* to *m*, there is not necessarily a function from the monad back to the free monad, such that the homomorphism forms an isomorphism with this function. For example, as mentioned by Swierstra [24], the list monad is not a free monad in the sense that the list monad is not isomorphic to an instance of the free monad. In order to prove statements about more complex effects Mona would have to be able to model monads like the list monad that were not free. Mona realized that she could not resort to structural equality as some values that were structurally equal in the original monad were not structurally equal in the the presentation using *Free f*. This divergence is due to the fact that more values inhabit the resulting representation using *Free* than inhabit the original monad, if the original monad is not a free monad. She observed, however, that she could use a custom equality for this kind of monad to solve this problem. When proving a statement in a setting with a monad that was not a free monad, she would use a custom equality instead of using structural equality.

---

[6] The monad *M* is equipped with the functions *ret* and *join*.
[7] Mona had to pass the type argument *M A* to the higher-ranked functional argument *f* used in the application of *join* because Coq was not able to infer it.

At first, defining this kind of custom equality felt like an involved task to Mona, because she had to compare two instances of the free monad that used a container extension as functor. However, Mona realized that she could define the equality by means of *induce* and an equality on the original monad. That is, in order to compare two instances of the free monad she could use the homomorphism to map these instances to the original monad and use structural equality on the monadic terms. This way she was able to model an effect that is not free. In a nutshell, Mona would interpret the *Free* values as their monadic representatives and compare the monadic values using structural equality.

In another round of poring over her manuscripts, Mona found an implementation by Verbruggen et al. [26] that uses polynomial functors to define generic data types in Coq. The encoding using polynomial functors represents data types as combinations of four primitive constructors: unit, identity, product, and sum. One disadvantage of this encoding is that many interesting data types cannot be represented: there is no possibility to represent function types at least not without running into problems concerning the strict positivity restriction, again. In contrast, Mona could reason about all monads *m* that are free monads in the sense that there exists a functor *f* such that *m* is isomorphic to *Free f*. Mona had the additional restriction that *f* had to be a container. For example, identity, maybe and error are free monads whose functors are containers.

If a monad *m* is not a free monad, that is, it is not isomorphic to some instance of *Free*, Mona could still reason about it using a custom equality as long as the functor can be modeled using a container. For example, the state monad is not a free monad but can be modeled using a container. Mona found implementations of more involved effects like state using the free monad in combination with a container.[8] Mona was quite astonished by the variety of monads she could model in her current setting. Then she once again considered the continuation monad that she failed to instantiate in the beginning of her journey. Indeed, it was still not possible to model the continuation monad using a container extension. It was a bit of a pity for Mona that she could not represent the continuation monad, nevertheless, for now, she was content with the variety of monads that she could represent with her encoding.

---

[8] For example, Mona found an implementation of the state monad in Agda using a free monad and containers (https://github.com/agda/agda-stdlib/blob/v2.4.0/README/Container/FreeMonad.agda (last accessed: 2019-01-28).

## 4 A Simple Proof

The next morning, Mona tackled the definition she struggled with in the beginning and that started this whole journey in the first place: a list with polymorphic elements and constructors with monadic components. Instead of using a polymorphic type constructor for the monadic effect, Mona parametrized the definition of a monadic list over a container $C_F$ and applied the type *Free $C_F$ A* to the arguments of the *cons* constructor.

Inductive *List* ($C_F$ : *Container F*) *A* :=
| *nil* : *List $C_F$ A*
| *cons* : *Free $C_F$ A* → *Free $C_F$* (*List $C_F$ A*) → *List $C_F$ A*.

Mona was quite happy with her implementation of a list with monadic effects. As a next step, she wanted to prove a property about a simple Haskell function. First, she had to translate her Haskell program into a monadic Coq program. Fortunately, the manuscript by Abel et al. [1] contained the formal definition of a transformation of Haskell programs into Agda programs.

Right now Mona did not care about a formal definition of the translation to Coq, but was more interested in reasoning about the manually transformed code in Coq. Nevertheless she realized that the translation from Haskell to Coq was an interesting topic in itself. For example, because programs in Coq had to terminate, the Coq termination checker had to be able to check that the monadic transformation of the program terminated. Mona wondered whether there was a transformation that, given that a non-monadic version was accepted by the termination checker, accepted the monadic translation as well.

In order to simplify the definition of monadic programs Mona defined the following smart constructors.

Definition *Nil A* : *Free $C_F$* (*List $C_F$ A*) := *pure nil*.

Definition *Cons A* (*fx* : *Free $C_F$ A*) (*fxs* : *Free $C_F$* (*List $C_F$ A*)) : *Free $C_F$* (*List $C_F$ A*) :=
  *pure* (*cons fx fxs*).

With these smart constructors and the monadic bind operator >>= at hand, Mona was ready to define functions on *List*s. As a simple example Mona defined list concatenation.

Fixpoint *append' A* (*xs*: *List $C_F$ A*) (*fys*: *Free $C_F$* (*List $C_F$ A*)) : *Free $C_F$* (*List $C_F$ A*) :=
  match *xs* with
  | *nil* ⇒ *fys*
  | *cons fz fzs* ⇒ *Cons fz* (*fzs* >>= λ *zs* ⇒ *append' zs fys*)
  end.

Definition *append A* (*fxs fys*: *Free $C_F$* (*List $C_F$ A*)) : *Free $C_F$* (*List $C_F$ A*) :=
  *fxs* >>= λ *xs* ⇒ *append' xs fys*.

*If you are keen to know why Mona had to define append using a helper function, you can follow Mona to get some insights about nested recursive function definitions in Coq in section A.2. If you are already familiar with nested recursive function definitions in Coq or simply do not need more details for now, just read on.*

**Induction Principle**

Finally, Mona had all the ingredients at hand to prove properties about Haskell functions. As she had read the manuscript by Abel et al. [1] she wanted to prove the associativity of *append* as well. In accordance with the manuscript, she started to argue about a total world, that is, using the identity monad.

A classical proof of the associativity of append used structural induction over the argument list. In Coq, a proof by structural induction used a lemma called the induction principle that is automatically generated for each data type. However, when Mona tried to apply the induction principle for the monadic list data type *List*, it did not work as expected at first.

In a nutshell, Mona needed to define a custom induction principle for the free monad data type *Free* as well as for *List*, the list data type with monadic components. One might wonder why she needed an induction principle for the data type *Free* at all. In the definition of the free data type at the beginning of section 3, the *impure* constructor builds a stack of applications of the functor. To reason about this stack of functor applications an induction principle was required.

*If you want to know why Mona failed to apply the induction principle, please read about all the insights she gained on induction principles in section A.3. If you already know all about nested inductive type definitions and their induction principles or do not need more details for now, just read on.*

With the definitions of induction principles named *Free_Ind* and *List_Ind*, respectively, at hand, Mona was eager to try her proof of the associativity of append again.

**Identity Monad**

Working with Coq opened a whole new, magical world to her. Proofs in Coq consisted of a sequence of spells called tactics. As Mona was always eager for knowledge, she read several spell books and tried to write a proof of the associativity of *append*. All the following lemmas that Mona wanted to prove were quantified over a type, so she wrote a side-note that *A* was an arbitrary type and used *A* as an implicit parameter.

Mona started with an auxiliary lemma for *append'*. The proof of the auxiliary lemma was mostly straightforward and similar to a proof for ordinary lists. In case of an empty list the statement was trivially true and in case of a non-empty list the statement followed from the induction hypothesis. In order to apply the induction hypothesis Mona used the custom tactic *simplify*. The shortcut *simplify H* as *IH* simplified a hypothesis *H* generated by *List_Ind* and introduced the required hypothesis under the name *IH*. This slight transformation was necessary due to the custom induction principle used for *List*.

```
Lemma append'_assoc_Id : ∀ (xs : List C_Zero A) (fys fzs : Free C_Zero (List C_Zero A)),
    append' xs (append fys fzs) = append (append' xs fys) fzs.
Proof.
  (* Let xs, fys, fzs be arbitrary. *)
  intros xs fys fzs.
  (* Perform induction over the List structure xs. *)
  induction xs using List_Ind.
  − (* Base case: xs = nil *)
    reflexivity.
  − (* Inductive case: xs = cons fx fxs with induction hypothesis H
        Perform induction over the Free structure fxs. *)
    induction fxs using Free_Ind.
    + (* Base case: fxs = pure x
          Simplify and use induction hypothesis IH. *)
      simpl. simplify H as IH. rewrite IH. reflexivity.
    + (* Inductive case: fxs = impure (ext s pf) with s of type Shape_Zero *)
      destruct s.
Qed.
```

There was an odd-looking assumption in the inductive case for *Free*: a value *s* of type *Shape_Zero*. That is, she had a value of a type that did not have any inhabitants. Mona realized that *s* was a false assumption; she simply had to call that bluff! The tactic destruct made a case distinction on its argument; in this case it made Coq realize that the value *s* cannot exist. This behavior actually started to make sense to Mona. Using *Zero* meant that she was currently proving something in a total setting. In a total setting, nothing can go wrong, so there actually was no *impure* case.

As a next step Mona proved the actual lemma that stated the associativity of *append* as follows.

```
Lemma append_assoc_Id : ∀ (fxs fys fzs : Free C_Zero (List C_Zero A)),
    append fxs (append fys fzs) = append (append fxs fys) fzs.
Proof.
  (* Let fxs, fys, fzs be arbitrary. *)
  intros fxs fys fzs.
  (* Perform induction over the Free structure of fxs. *)
  induction fxs using Free_Ind.
  − (* Base case: fxs = pure x
        Simplify and use auxiliary lemma. *)
    simpl. apply append'_assoc_Id.
  − (* Inductive case: fxs = impure (ext s pf) with s of type Shape_Zero *)
    destruct s.
Qed.
```

In the base case of the induction over the *Free* structure Mona simply used the tactic apply to use the lemma she had proven before. In the inductive case she again had a value that did not exist and, therefore, got rid of it by using destruct.

**One Monad to Prove Them All**

Mona was quite content and thought that her proof was reasonably simple, however, she found it still slightly more complex than the original proof by Abel, Benke, Bove, Hughes, and Norell. For example, Mona had to handle the *impure* case explicitly, which does not exist when using the actual identity monad.

**Maybe Monad**

As Mona was happy with the proof in the total setting, as a next step she considered a setting with partial values, that is, using *One* as container instead of *Zero*. She expected the proofs to be similar to the ones before, as one should be able to handle the *pure* cases in the same way. And, indeed, the proofs were nearly identical and only differed in the inductive cases for *Free*.

Lemma *append'_assoc$_{Maybe}$* : ∀ (*xs* : *List C$_{One}$ A*) (*fys fzs* : *Free C$_{One}$* (*List C$_{One}$ A*)),
    *append' xs* (*append fys fzs*) = *append* (*append' xs fys*) *fzs*.
Proof.
  . . .
    + (* Inductive case: *fxs* = *impure* (*ext s pf*)
        Simplify, drop *Cons*, *impure*, and *ext s* on both sides. *)
      simpl. do 3 apply f_equal.
      (* Use functional extensionality and case distinction over *p* : *Pos$_{One}$ s* *)
      extensionality *p*. destruct *p*.
Qed.

Lemma *append_assoc$_{Maybe}$* : ∀ (*fxs fys fzs* : *Free C$_{One}$* (*List C$_{One}$ A*)),
    *append fxs* (*append fys fzs*) = *append* (*append fxs fys*) *fzs*.
Proof.
  . . .
  − (* Inductive case: *fxs* = *impure* (*ext s pf*)
      Simplify, drop *impure*, and *ext s* on both sides. *)
    simpl. do 2 apply f_equal.
    (* Use functional extensionality and case distinction over *p* : *Pos$_{One}$ s*. *)
    extensionality *p*. destruct *p*.
Qed.

The proofs actually had to consider a potential value in the *impure* cases, because the shape *s* of the underlying container *One* is *Shape$_{One}$*, a type that has exactly one inhabitant. Thus, Mona could not take the easy way out relying on a false assumption again. When she took a second close look, Mona realized that there was still something odd about the given assumptions. Mona got rid of the constructors *Cons*, *impure* and *ext* that both sides of the equality had in common by using f_equal[9] thrice. After using f_equal Mona ended up with two functions. Both functions took an argument of type *Pos$_{One}$ s*, again a type without inhabitants.

As Mona came from the kingdom of Haskell, she was used to the rule of functional extensionality. That is, if two functions behave the same for all possible arguments, the functions are considered equal. In Coq, two functions are equal with respect to the equality = if these functions evaluate to the same value or are defined in the exact same way, expression by expression, character by character modulo renaming. If two functions just behaved the same for all possible arguments, Mona needed the axiom of functional extensionality in order to use this property in Coq. Fortunately, this axiom is consistent with the Calculus of Inductive Constructions, that is, with the lambda calculus that is the formal basis of Coq. After applying the rule of functional extensionality Mona had to show that the two functions behaved the same for all possible inputs. However, as the argument type of such an input did not contain any inhabitants, there were no arguments and therefore, the statement was trivially valid. Mona used destruct again to prove that there were no possible values for *p*.

---

[9] She read that the spell f_equal was sometimes also called congruence or cong in other magical kingdoms.

Mona understood that she was able to finish the proof with destruct because the free monad that represented the maybe monad was quite simple. The *impure* case represented the constructor nothing and this constructor did not contain any value.

**Monad-Generic Proof**

Mona was very satisfied with her solution. Up until now, she was able to prove statements about Haskell programs when only considering total values and was also able to consider partial values. She consulted the manuscript by Abel et al. [1] again and read the following paragraph.

> The reader may well wonder why we prove the same property twice, for two different monads — why not just prove it once-and-for-all, for any monad? While this may seem attractive in principle, it turns out to be much more difficult in practice. [...] Since Agda does not know how to compute with a general monad it will not be able to simplify the type of the properties by performing reduction steps. Thus, the only thing we can do to prove properties is to use the monad laws explicitly. Although it is possible to prove properties in this way, those proofs are both more difficult to perform and to understand, and much longer than the ones we presented above.

The scales fell from Mona's eyes. These statements did not hold in her setting. When using a free monad underneath, a generic proof was not that difficult anymore. If Mona proved a statement for the free monad for an arbitrary container, the statement would hold for a whole class of monads — with the minor restriction that the monad can be represented by a container-based instance of the free monad. As this possibility was more than Mona had originally hoped for, she immediately began to write the following monad-generic proofs.

Lemma *append'_assoc* : ∀ (*xs* : *List $C_F$ A*) (*fys fzs* : *Free $C_F$* (*List $C_F$ A*)),
    *append' xs* (*append fys fzs*) = *append* (*append' xs fys*) *fzs*.
Proof.
  ...
    + (* Inductive case: *fxs* = *impure* (*ext s pf*) with induction hypothesis *H* *)
      simpl. do 3 apply f_equal. extensionality *p*.
      *simplify H* as *IH*. apply *IH*.
Qed.

```
Lemma append_assoc : ∀ (fxs fys fzs : Free C_F (List C_F A)),
    append fxs (append fys fzs) = append (append fxs fys) fzs.
Proof.
  ...
  – (* Inductive case: fxs = impure (ext s pf) with induction hypothesis IH *)
    simpl. do 2 apply f_equal. extensionality p.
    apply IH.
Qed.
```

Again these proofs basically followed the same structure. Besides the generic parameter $C_F$ for the container, once again, the only crucial differences occurred in the inductive cases for *Free*. In general, Mona could not hope for false assumptions to complete the proofs. A more complex effect, for instance non-determinism, could actually have a recursive occurrence in the *impure* case. In this case the induction principle for *Free* came in handy. Mona used the induction hypothesis that said that the given statement already holds for all values that resulted from applying the position function *pf*. More precisely, the induction hypothesis provided the following statement, which states that the statement holds for lists resulting from *pf p* for all possible positions *p*.

∀ *p* : *Pos s*, *append* (*pf p*) (*append fys fzs*) = *append* (*append* (*pf p*) *fys*) *fzs*

After using functional extensionality she was able to simply apply this hypothesis by using `apply IH`. In contrast, in the case of *Zero* there was no *impure* case and in the case of *One* the *impure* case did not contain any values. Thus, Mona did not need to use the induction hypothesis for these concrete instances.

As hard days of working through Coq tactics and these proofs passed, Mona was very pleased with her current approach. She was even more than thrilled that using the free monad gave rise to generic proofs that hold for a whole class of monads. Furthermore, she observed that monad-generic proofs were not difficult at all. The case for the *pure* constructor was not difficult because it is closely connected to the corresponding case without effects. Even the proof for the *impure* constructor was reasonable simple. In the *impure* case Mona always had to prove — with an induction hypothesis at hand — that the statement holds for all possible positions.

## 5 A Case Study

Proving a more or less trivial property for a single list function was not exactly setting the world on fire. Thus, Mona was eager to try her approach to prove the correctness of actual Haskell code. Since she was helping out with some classes at the department she stayed at, she seized the opportunity to teach her approach in class. The main topic of the class was to compare two implementations of queues. There was a naive implementation using a single list and a more advanced implementation that used two lists to improve the performance. In class they started with an implementation in Haskell and discussed that both concrete implementations are interchangeable with one another; the reasoning was backed up by property-based testing via QuickCheck [14]. Mona thought that the implementations would be a particularly good example, because they contained total as well as partial functions[10]. That is, some of the statements were only valid in the context of the maybe monad, while most statements held in the total as well as in the partial setting.

```
type Queue a = [a]                type QueueI a = ([a],[a])       prop_isEmpty qi =
                                                                    invariant qi ==>
empty :: Queue a                  emptyI :: QueueI a                  isEmptyI qi == isEmpty (toQueue qi)
empty = []                        emptyI = ([],[])
                                                                  prop_add x qi =
isEmpty :: Queue a -> Bool        isEmptyI :: QueueI a -> Bool      toQueue (addI x qi) == add x (toQueue qi)
isEmpty q = null q                isEmptyI (f,b) = null f
                                                                  prop_front qi =
front :: Queue a -> a             frontI :: QueueI a -> a           invariant qi && not (isEmptyI qi) ==>
front (x:q) = x                   frontI (x:f,b) = x                  frontI qi == front (toQueue qi)

add :: a -> Queue a -> Queue a    addI :: a -> QueueI a -> QueueI a
add x q = q ++ [x]                addI x (f,b) = flipQ (f,x:b)

                                  flipQ :: QueueI a -> QueueI a
                                  flipQ ([],b) = (reverse b,[])
                                  flipQ q      = q
```

■ **Scroll 1** A queue implementation based on a single list (left) and one based on two lists (middle) as well as QuickCheck properties relating these implementations (right).

Scroll 1 shows two different implementations of queues in Haskell. The implementation to the left uses a single list to model the queue. The implementation in the middle uses two lists in order to provide a more efficient implementation of the function add. Mona defined the following function toQueue to relate these two implementations.

```
toQueue :: QueueI a -> Queue a
toQueue (f,b) = f ++ reverse b
```

In order to prove statements about the functions shown in Scroll 1 in Coq, Mona started to transform the data types Queue and QueueI to corresponding monadic versions. For the definition of QueueI, she additionally needed a monadic version of pairs. The data type Queue is just a type synonym for a list, which can be defined in Coq as a type synonym as well.

   Inductive *Pair* ($C_F$ : *Container F*) *A B* := *pair* : *Free $C_F$ A* → *Free $C_F$ B* → *Pair $C_F$ A B*.

   Definition *QueueI* ($C_F$ : *Container F*) *A* := *Pair $C_F$* (*List $C_F$ A*) (*List $C_F$ A*).

   Definition *Queue* ($C_F$ : *Container F*) *A* := *List $C_F$ A*.

The functions *front* and *frontI* were partially defined because the cases for the empty list were missing in these definitions. That is, when modeling these functions in Coq, Mona had to use the maybe instance. Mona only considered the implementation of *front* as the definition of *frontI* followed the same idea. Moreover, she defined the Coq version of toQueue to relate both implementations. The transformations of all other functions to Coq were straightforward and followed the approach used for *append*. Mona used the constant *Nothing* as defined in section 3 to simplify the definitions.

---

[10] In fact, Mona had also read about this example in the manuscript by Abel et al. [1].

Definition *front A* (*fq* : *Free C<sub>One</sub>* (*Queue C<sub>One</sub> A*)) : *Free C<sub>One</sub> A* :=
  *fq* >>= λ *q* ⇒ match *q* with
         | *nil* ⇒ *Nothing*
         | *cons x _* ⇒ *x*
         end.

Definition *toQueue A* (*fqi*: *Free C<sub>F</sub>* (*QueueI C<sub>F</sub> A*)) : *Free C<sub>F</sub>* (*Queue C<sub>F</sub> A*) :=
  *fqi* >>= λ '(*pair ff fb*) ⇒ *append ff* (*reverse fb*).

In class Mona discussed the QuickCheck properties given in the right column in Scroll 1. The operator ==> took a boolean expression that was checked as a precondition for the test case. The function invariant stated an invariant about the QueueI implementation, namely, that the back end of the queue would be empty whenever the front end was empty. In other words, either the back end was empty or the front end was not empty. This property was checked by the following predicate on queues.

```
invariant :: QueueI a -> Bool
invariant (f,b) = null b || not (null f)
```

When writing test cases in QuickCheck, the programmer usually considers total values only. Therefore, Mona extended the invariant with the requirement that all queues that are checked are total using the proposition *total_list*. While Mona used the maybe instance as model, she did not define *total_list* for the special case of the maybe monad, but generalized it to an arbitrary effect. That is, while she still named the predicate *total_list*, it rather checked whether the list was effect-free.

Inductive *total_list A* : *Free C<sub>F</sub>* (*List C<sub>F</sub> A*) → Prop :=
| *total_nil* : *total_list* (*pure nil*)
| *total_cons* : ∀ *fx fxs*, *total_list fxs* → *total_list* (*pure* (*cons fx fxs*)).

The definition needed to distinguish between two cases. First, a defined empty list was total, that is, *pure nil* was total. Second, a defined non-empty list *pure* (*cons fx fxs*) was total if the tail *fxs* was total as well. The proposition did not check whether the element *fx* was defined, that is, whether it had the form *pure x*.

Since Mona wanted to work with properties about *QueueIs*, she extended the definition to check both list components of a queue. The functions *fst* and *snd* projected to the components of a pair lifted into the free monad.

Definition *total_queue A* (*fqi* : *Free C<sub>F</sub>* (*QueueI C<sub>F</sub> A*)) : Prop :=
  *total_list* (*fst fqi*) ∧ *total_list* (*snd fqi*).

In order to transform the Haskell function invariant to Coq, Mona implemented monadic liftings of the functions *null* and the boolean combinators *not*, (||) as well as (&&). Now Mona was ready to implement the invariant.

Definition *invariant A* (*fqi* : *Free C<sub>F</sub>* (*QueueI C<sub>F</sub> A*)) : *Free C<sub>F</sub> bool* :=
  *null* (*snd fqi*) || *not* (*null* (*fst fqi*)).

Since the QuickCheck properties all used premises whose type is *Free C<sub>F</sub> bool*, Mona defined a smart constructor (⟹) to lift these *Free C<sub>F</sub> bool* values into properties.

First Mona showed the class a proof of a property that was specific to the maybe monad. The property was specific to the maybe monad because it made use of the partial function *front*. The property stated that *frontI* yields the same result as a combination of *front* and *toQueue* under the precondition that the queue fulfilled the invariant and was not empty. Mona explained the students that all the following proofs assumed *A* to be an arbitrary type, that is, *A* was used but not explicitly introduced in the code she wrote.

    Lemma *prop_front* : ∀ (*fqi* : *Free* $C_{One}$ (*QueueI* $C_{One}$ *A*)),
        *total_queue fqi* →
        *invariant fqi* && *not* (*isEmptyI fqi*) ⟹ *frontI fqi* = *front* (*toQueue fqi*).

In contrast to the maybe-specific properties, the properties *prop_isEmpty* and *prop_add* held independently of the considered monad and could be proven once and for all. Trivially, this monad-generic proof also implied that the same property held in the special case of *Free* $C_{One}$. Mona wrote a monad-generic proof of the following property.

    Lemma *prop_isEmpty* : ∀ (*fqi* : *Free* $C_F$ (*QueueI* $C_F$ *A*)),
        *total_queue fqi* →
        *invariant fqi* ⟹ *isEmptyI fqi* = *isEmpty* (*toQueue fqi*).

Some time went by while Mona showed the class how to prove one of the QuickCheck properties after another. At last, she proved the property *prop_add*.

    Lemma *prop_add* : ∀ (*fa* : *Free* $C_F$ *A*) (*fqi* : *Free* $C_F$ (*QueueI* $C_F$ *A*)),
        *toQueue* (*addI fa fqi*) = *add fa* (*toQueue fqi*).

After all proofs were done, one of Mona's students complained about the definition of *front*. For the sake of the user, the function *front* should yield an error that explained the problem instead of simply raising a generic non-exhaustive patterns error. As her student's wish was her command, she implemented the following variant of *front* in Haskell.

```
front :: Queue a -> a
front (x:f,b) = x
front _ = error "front: empty queue"
```

One of the other students now complained that they had to do all the nice proofs again but using the functor *Const* instead of *One*. Mona appeased them by pointing out that they only had to do the maybe-specific proofs like *prop_front* again, but not all the monad-generic proofs like *prop_isEmpty* and *prop_add*. That is, they only had to redo proofs that talked about function definitions that actually changed. In fact it all fit together quite nicely. For example, as the predicate *total_list* checked whether a list was effect-free, Mona reused it in the case of the error effect and it specified exactly what one was looking for: a list without errors.

## 6 The End

Happy with her results, Mona headed back home to the kingdom of Haskell the next week. She had learned a lot about proving properties of Haskell programs. Well-known concepts like monadic transformations of data types, containers and free monads were the foundations of her solution. The combination of all of these ideas led to a new approach to model and prove properties about effectful programs in Coq. Even more importantly, while Abel et al. [1] thought it difficult to prove monad-generic statements, Mona had discovered that it was not that difficult after all. She was able to prove statements using an arbitrary effect represented using the free monad, whose functor is a container.

Mona really looked forward to continuing to pursue her PhD because there were so many questions she would like to have an answer to. For now she only considered the effect of partiality, while her proofs held for arbitrary effects. For example, using her approach she could model lazy tracing as provided by the function *trace* by means of a simple reader monad. The best news about her current setup was that all the monad-generic proofs would still hold in this setting.

Besides extending her simple model of Haskell with more advanced effects, Mona planned to use her setting to model other programming languages. For example, a functional logic programming language like Curry can be modeled by the same transformation as shown by Fischer et al. [13]. This model uses a non-determinism monad instead of the partiality monad. Mona also planned to investigate modeling a probabilistic programming language as she read that functional logic and probabilistic programming are closely related in a manuscript by Dylus et al. [12].

When Mona looked into effects like non-determinism, she observed that there were some additional requirements. While the free monad extends a functor $f$ with an additional structure to define the functions *ret* and *bind*, the free monad provides exactly this structure only. However, some monads provide more structure than given by a free monad. Mona thought about the list monad again, which was not a free monad and needed a custom equality in her setting as discussed in section 3. In order to prove statements that rely on additional properties of a concrete monad, Mona could not resort to structural equality, she would have to define a custom equality. However, defining custom equalities for free monads by means of the corresponding monad homomorphism is straightforward. Furthermore, monad-generic proofs can also be reused in a setting with a custom equality because terms that are definitionally equal are also equal with respect to a custom equality.

Thinking about different effects raised the question of whether the translation of programs had to be adapted. Mona's current translation was based on the optimized version by Abel et al. [1], where partiality was the only effect that could occur. In this case a function $f$ of type $\tau \to \tau'$ could be modeled by a Coq function of type $M\ \tau \to M\ \tau'$. However, in a more general setting, Mona had to use a function of type $M\ (M\ \tau \to M\ \tau')$ instead. For example, if *seq* is available, one can observe whether a function $f$ is defined as *undefined* or as $\lambda\ x \Rightarrow undefined$. For instance, the expression *seq f* 42 fails in the former case, but yields 42 in the latter case.

Using function types of the form $M\ (M\ \tau \to M\ \tau')$ is in particular important, when modeling a functional logic language. For example, some free theorems fail in the context of functional logic languages due to the difference between a non–deterministic choice of two functions and a function that non-deterministically yields two results as shown by Christiansen et al. [7] and elaborated by Mehner et al. [22]. Mona wanted to use her framework for a variety of effects, thus, the optimized translation of function types used by Abel et al. [1] was not applicable anymore.

Finally, Mona wondered about other possibilities to model the monadic data types she wanted to define. One possible alternative was to use a more direct modeling of strictly positive monads. Currently, she used a free monad whose functor is a container. Instead she could as well use only a container. For example, the identity and the maybe monad can be modeled as containers. That is, she could use container monads as discussed by Uustalu [25], which model containers that provide a monadic structure. This way she could even model the list monad without using a custom equality. However, when using container monads directly Mona could only reason about monadic programs using monad laws. Abel et al. [1] did not consider monad-generic proofs because they had to resort to monad laws. In contrast, Mona's current approach using free monads enabled Coq to compute with a general monad and simplify properties by reduction. Another alternative was to use freer monads introduced by Kiselyov et al. [17], whose definition did not need a representation of striclty positive types using containers at all. There were already several encodings to model different kinds of monadic effects using freer monads: McBride [21] defines a *General* monad to model general recursion as effect, Letan et al. [19] use the *Program* monad initially presented in the operational package known from Haskell[11] to reason about a small imperative language, and Koh et al. [18] identify *interaction trees* as a suitable tool to verify functional correctness of a server implemented in C.

The case study about queues only involved two data structures: lists and pairs. Lifting data types into their monadic counterparts follows a general scheme. Mona found a manuscript by Atkey et al. [4] that showed how to benefit from describing inductive types interleaved with effects as initial f-and-m-algebras. Their description of data types interleaved with effects was similar to the monadic lifting that Mona pursued. The benefit of using *initial f-and-m-algebras* came from the clear separation of monadic effects described by the *m* and the pure data structure represented by a functor *f*. This observation reminded Mona of her own proofs about the associativity of *append* as well as other proofs using induction. There were always two cases to consider: the pure case and the impure case, where the pure cases was then again divided into the different constructors of the data type involved. Mona was eager to find out if her approach can also benefit from the abstractions used in the work of Atkey and Johann.

As Mona knew that a successful PhD student would interest her peers in her results, she started to spread the word about her work. In order to practice her writing skills, she also wrote a short story about her journey through the land of functional programming. Mona continued to pursue her PhD and lived happily ever after.

**Acknowledgements**   We thank the Coq community for being supportive and answering questions about Coq. Furthermore, we are thankful for the remarks and suggestions from the anonymous reviewers as well as the valuable comments of Jan Bracker, Sebastian Fischer, and Nikita Danilenko on a draft version of this paper.

---

[11] http://hackage.haskell.org/package/operational (last accessed: 2019-01-28).

One Monad to Prove Them All## References

[1] Andreas Abel, Marcin Benke, Ana Bove, John Hughes, and Ulf Norell. "Verifying Haskell Programs Using Constructive Type Theory". In: *Proceedings of the 2005 ACM SIGPLAN Workshop on Haskell*. ACM, 2005, pages 62–73. DOI: 10.1145/1088348.1088355.

[2] Michael Abott, Thorsten Altenkirch, and Neil Ghani. "Categories of Containers". In: *Proceedings of Foundations of Software Science and Computation Structures*. Volume 2620. Lecture Notes in Computer Science. Springer-Verlag, 2003, pages 23–38. ISBN: 978-3-540-36576-1. DOI: 10.1007/3-540-36576-1.

[3] Thorsten Altenkirch, Conor McBride, and Peter Morris. "Generic Programming with Dependent Types". In: *International Spring School*. Volume 4719. Lecture Notes in Computer Science. Springer-Verlag, 2006, pages 209–257. ISBN: 978-3-540-76786-2. DOI: 10.1007/978-3-540-76786-2_4.

[4] Robert Atkey and Patricia Johann. "Interleaving Data and Effects". In: *Journal of Functional Programming* 25 (2015). DOI: 10.1017/S0956796815000209.

[5] Joachim Breitner, Antal Spector-Zabusky, Yao Li, Christine Rizkallah, John Wiegley, and Stephanie Weirich. "Ready, Set, Verify! Applying Hs-to-Coq to Real-World Haskell Code (Experience Report)". In: *Proceedings of the ACM on Programming Languages* ICFP (2018), 89:1–89:16. DOI: 10.1145/3236784.

[6] Adam Chlipala. *Certified Programming with Dependent Types*. MIT Press New York, 2011. ISBN: 978-0-262-02665-9.

[7] Jan Christiansen, Daniel Seidel, and Janis Voigtländer. "Free Theorems For Functional Logic Programs". In: *Proceedings of the 4th ACM SIGPLAN Workshop on Programming Languages Meets Program Verification*. ACM, 2010, pages 39–48. DOI: 10.1145/1707790.1707797.

[8] Thierry Coquand and Christine Paulin. "Inductively Defined Types". In: *Proceedings of the International Conference on Computer Logic*. Volume 417. Lecture Notes in Computer Science. Springer-Verlag, 1988, pages 50–66. ISBN: 978-3-540-46963-6. DOI: 10.1007/3-540-52335-9_47.

[9] Thierry Coquand, Aaron Stump, and Wojciech Jedynak. *Agda Mailing List*. Dec. 18, 2013. URL: https://lists.chalmers.se/pipermail/agda/2013/006189.html (visited on 2019-01-28).

[10] Nils Anders Danielsson, John Hughes, Patrik Jansson, and Jeremy Gibbons. "Fast and Loose Reasoning Is Morally Correct". In: *Conference Record of the 33rd ACM SIGPLAN-SIGACT Symposium on Principles of Programming Languages*. ACM, 2006, pages 206–217. DOI: 10.1145/1111320.1111056.

[11] Sandra Dylus. *ichistmeinname/free-proving: Revised Submission to <Programming>*. Dec. 2018. DOI: 10.5281/zenodo.2529451. URL: https://doi.org/10.5281/zenodo.2529451.
8:24

## A  A Trip to the Technical Corners of the Land of Dependently Typed Programming

### A.1  Definition of fold and bind

Mona wanted to translate the Haskell definition of the functions fold and bind on *Free* she found in the manuscript by Swierstra [24]. She started with fold as it looked like the easier task.

```
foldFree :: Functor f => (a -> b) -> (f b -> b) -> Free f a -> b
foldFree pur imp (Pure x) = pur x
foldFree pur imp (Impure fx) = imp (fmap (foldFree pur imp) fx)
```

In contrast to Swierstra, Mona worked on a container representation that she had to transform into the corresponding functor using *to* first in order to use *fmap*. However, instead of using the function *fmap* she could use a concrete mapping function *cmap* on *Ext* directly. This way she could get rid of the additional *Functor* constraint.

Mona defined the following map function for the container $C_F$, where *Shape* and *Pos* are the types associated with the container.

Definition *cmap* $A\ B\ (f : A \to B)\ (x : Ext\ Shape\ Pos\ A) : Ext\ Shape\ Pos\ B :=$
  match $x$ with
  | *ext s pf* $\Rightarrow$ *ext s* ($\lambda\ x \Rightarrow f\ (pf\ x)$)
  end.

Based on the definition of *cmap*, she implemented the following recursive function to fold *Free* expressions. As the function *imp* worked on arguments of type *F B*, Mona applied the function *to* on the argument of *imp* to transform the container representation into the concrete structure *F*.

Fixpoint *fold_free* $A\ B\ (pur : A \to B)\ (imp : F\ B \to B)\ (fx : Free\ C_F\ A) : B :=$
  match *fx* with
  | *pure x* $\Rightarrow$ *pur x*
  | *impure e* $\Rightarrow$ *imp* (*to* (*cmap* (*fold_free pur imp*) *e*))
  end.

Besides the special handling of the container construction, the translation was quite straightforward. Next up, Mona took another look at the Haskell definition of the bind operator for *Free*.

```
(»=) :: Functor f => Free f a -> (a -> Free f b) -> Free f b
Pure x »= f = f x
Impure fx »= f = Impure (fmap (»= f) fx)
```

For her translation to Coq, she did not need to use *fmap*, as above, but the auxiliary function *cmap*. Otherwise the two definitions looked quite alike.

Fixpoint *free_bind* $A\ B\ (fx : Free\ C_F\ A)\ (f : A \to Free\ C_F\ B) : Free\ C_F\ B :=$
  match *fx* with
  | *pure x* $\Rightarrow f\ x$
  | *impure e* $\Rightarrow$ *impure* (*cmap* ($\lambda\ x \Rightarrow$ *free_bind x f*) *e*)
  end.

Only later, Mona realized that the definition caused problems. When she tried to define recursive functions that use *free_bind*, Coq's termination checker nagged about most of her programs.

As Mona studied a lot of Coq code during her journey, she remembered that recursive functions were often defined using so-called Sections. For example, the standard library for lists in Coq follows the same approach to define functions like *map*, *fold_right* and *fold_left*. She changed her code as follows.

Section *fbind*.

    Variable *A B*: Type.
    Variable *f*: *A* → *Free $C_F$ B*.

    Fixpoint *free_bind'* (*ffA*: *Free $C_F$ A*) :=
      match *ffA* with
      | *pure x* ⇒ *f x*
      | *impure* (*ext s pf*) ⇒ *impure* (*ext s* (λ *p* ⇒ *free_bind'* (*pf p*)))
      end.

End *fbind*.

At the start of the section Mona introduced all variables annotated with their types that would be used throughout the section. All definitions in this section take these variables as additional arguments when used outside of the section. That is, within the section the function *free_bind'* had the type *Free $C_F$ A* → *Free $C_F$ B*, whereas it had the type (*A* → *Free $C_F$ B*) → *Free $C_F$ A* → *Free $C_F$ B* outside of the section. That is, within the section the only remaining argument of *free_bind'* was the inductive argument that was decreasing: the *Free* structure *ffA*. Mona read that it was crucial to define the function *f* as a section variable in order to signalize Coq's termination checker that the function *f* never changed in any recursive call. Furthermore, Mona defined the operator >>= based on *free_bind'*, but with arguments swapped.

### A.2 Nested Recursive Function Definitions

As the *Free* data type was a monad, Mona wanted to define *append* quite naturally using monadic notation as follows.

    Fail Fixpoint *append A* (*fxs fys* : *Free $C_F$* (*List $C_F$ A*)) : *Free $C_F$* (*List $C_F$ A*) :=
      *fxs* >>= λ *xs* ⇒ match *xs* with
                         | *nil* ⇒ *fys*
                         | *cons fz fzs* ⇒ *Cons fz* (*append fzs fys*)
                         end.

Unfortunately, Coq did not accept this definition of *append* because the termination checker was not able to guess the decreasing argument. However, Coq's termination checker accepted the following definition that was based on a helper function.

    Fixpoint *append' A* (*xs*: *List $C_F$ A*) (*fys*: *Free $C_F$* (*List $C_F$ A*)) : *Free $C_F$* (*List $C_F$ A*) :=
      match *xs* with
      | *nil* ⇒ *fys*
      | *cons fz fzs* ⇒ *Cons fz* (*fzs* >>= λ *zs* ⇒ *append' zs fys*)

end.

Definition *append* $A$ (*fxs*: *Free* $C_F$ (*List* $C_F$ $A$)) (*fys*: *Free* $C_F$ (*List* $C_F$ $A$)) : *Free* $C_F$ (*List* $C_F$ $A$) :=
  *fxs* >>= $\lambda$ *xs* $\Rightarrow$ *append'* *xs* *fys*.

Mona discovered that she could please Coq's termination checker by splitting the definition into two parts: a recursive function *append'* that had an argument of type *List A* and a non-recursive function *append* that called *append'* after unwrapping the monadic effect.

The main problem with Mona's original definition was that *List A* was a nested inductive type, *Free* (*List A*) occurred nested in *List A*. That is, to define a recursive function, Mona had to use nested recursion in the definition of *append'* as she had read about in the manuscript by Chlipala [6]. More precisely, the function *append'* was nested recursive because >>= was recursive and used nested in *append*. In the second branch of *append'* she had to inline the definition of *append* in order to resemble the nested recursive structure of the *List A* data type.

### A.3 Induction Principles

In order to understand what exactly was going on, Mona took some time for getting a better understanding of some technicalities of the programming language Coq. She started to look at the induction principle for *nat* in Coq.[12]

*nat_ind* : $\forall$ $P$ : *nat* $\rightarrow$ Prop,
  $P$ $O$
  $\rightarrow$ ($\forall$ $m$ : *nat*, $P$ $m$ $\rightarrow$ $P$ ($S$ $m$))
  $\rightarrow$ $\forall$ $n$ : *nat*, $P$ $n$

A proposition $P$ $n$ holds for all $n$ : *nat*, if $P$ holds for $O$ and, for all $m$ : *nat*, $P$ holds for $S$ $m$ given that it already holds for $m$. For all $m$ : *nat* the proposition $P$ $m$ is called the induction hypothesis.

Next, Mona took a look at the induction principle that was generated for the monadic list type *List*[13].

*List_ind* : $\forall$ ($A$ : Type) ($P$ : *List* $C_F$ $A$ $\rightarrow$ Prop),
  $P$ (*nil* $C_F$ $A$)
  $\rightarrow$ ($\forall$ (*fx* : *Free* $C_F$ $A$) (*fxs* : *Free* $C_F$ (*List* $C_F$ $A$)), $P$ (*cons fx fxs*))
  $\rightarrow$ $\forall$ (*l* : *List* $C_F$ $A$), $P$ $l$

This induction principle was supposed to be quite similar to the induction principle for Peano numbers. However, while there was a base case for the empty list, that is $P$ (*nil* $C_F$ $A$), the induction principle demanded that the proposition held for the non-empty list, that is $P$ (*cons fx fxs*), without having an induction hypothesis for *fxs*.

In order to prove a statement about *List*, Mona also needed an induction principle for *Free*. When Mona checked the induction principle for *Free*, it showed the same problem as the induction principle for *List*. She would have to prove $P$ (*impure e*) without knowing anything about *e*.

---

[12] The command *Check nat_ind* prints the type of the induction principle *nat_ind*.
[13] Coq generates an induction principle called *List_ind* for a data type called *List*.

*Free_ind* : $\forall$ (*A* : Type) (*P* : *Free $C_F$ A* $\to$ Prop),
  ($\forall$ *x* : *A*, *P* (*pure x*))
  $\to$ ($\forall$ *e* : *Ext Shape Pos* (*Free $C_F$ A*), *P* (*impure e*))
  $\to$ $\forall$ (*fx* : *Free $C_F$ A*), *P fx*

After some research Mona found a chapter in Chlipala's manuscript that explained how to define induction principles that fit one's purpose. The chapter also stated that the induction principles of nested data types generated by Coq are too weak (most of the time). Now all Mona had to do was to define her own induction principle for *List* and *Free*. Since *Free* is used nested in *List*, Mona started with *Free*.

Mona modified the type of *Free_ind* in order to fit her purposes and named the new induction principle *Free_Ind*. She only had to modify the last argument of *Free_ind* that handled the *impure* case. In the *impure* case, she had a value *e* : *Ext Shape Pos* (*Free A*), that is, a container extension with two components *s* : *Shape* and *pf* : *Pos s* $\to$ *Free A*. Intuitively the *impure* constructor contains a container extension of values of type *Free A*. Obviously, Mona needed the proposition *P* to hold for all elements of the container extension. Therefore, she added the requirement that every *Free A* that can be produced by the position function *pf* : *Pos s* $\to$ *Free A* satisfied proposition *P*. If that was the case, then *P* held for *impure* (*ext s pf*). Mona made the following changes.

*Free_Ind* : . . .
  . . .
  $\to$ ($\forall$ (*s* : *Shape*) (*pf* : *Pos s* $\to$ *Free $C_F$ A*), ($\forall$ *p*, *P* (*pf p*)) $\to$ *P* (*impure* (*ext s pf*)))
  . . .

Finally, Mona had to provide a function of the type of *Free_Ind*. Mona thought that the puzzle pieces fitted together quite naturally when she defined the following function via pattern matching on *fx* : *Free A*.

Fixpoint *Free_Ind A* (*P* : *Free $C_F$ A* $\to$ Prop)
  (*pur* : $\forall$ (*x* : *A*), *P* (*pure x*))
  (*imp* : $\forall$ (*s* : *Shape*) (*pf* : *Pos s* $\to$ *Free $C_F$ A*), ($\forall$ *p*, *P* (*pf p*)) $\to$ *P* (*impure* (*ext s pf*)))
  (*fx* : *Free $C_F$ A*) : *P fx* :=
  match *fx* with
  | *pure x* $\Rightarrow$ *pur x*
  | *impure* (*ext s pf*) $\Rightarrow$ *imp s pf* ($\lambda$ *p* : *Pos s* $\Rightarrow$ *Free_Ind P pur imp* (*pf p*))
  end.

In case of *pure x*, Mona had to provide a term of type *P* (*pure x*) and constructed it by applying the function *pur* to *x*. For *impure* (*ext s pf*), the function *imp* came in handy. As first two arguments Mona used the arguments *s* and *pf*; as third argument she needed a proof that the proposition *P* already held for *pf p* for all appropriate positions *p*, that is, an expression of type (*p* : *Pos s*) $\to$ *P* (*pf p*). The induction principle *Free_Ind* had type (*f x* : *Free A*) $\to$ *P fx* and, thus, matched the required type when Mona applied *pf* to a given position *p*. The other arguments used in the application of *Free_Ind* are the proposition *P* as well as the hypotheses *pur* and *imp* that remained unchanged.

Based on her new knowledge about the definition of induction principles and the example in Chlipala's manuscript [6], Mona also successfully defined an induction principle *List_Ind* for the data type *List*. Mona used induction for ordinary lists all the time. In the base case, the statement had to be proven for the empty list. For a non-empty list *cons x xs* there was a hypothesis in place that said that the statement already hold for *xs*. In the definition of *List* both arguments of *Cons* were wrapped in *Free*, such that Coq did not generate any hypotheses for the remaining list in the induction principle presented earlier.

*List_ind* : ∀ (A : Type) (P : List $C_F$ A → Prop),
  P (nil $C_F$ A)
  → (∀ (fx : Free $C_F$ A) (fxs : Free $C_F$ (List $C_F$ A)), P (cons fx fxs))
  → ∀ (l : List $C_F$ A), P l

The same problem occurs when an induction principle is generated for an arbitrary data structure *DT* that contains a list of *DT* as argument in one of its constructors. In the case of nested lists, an additional property is commonly used to lift propositions for *A* to lists of *A*, that is, to state that the property holds for all elements of a list. Mona applied this idea and defined the following proposition *ForFree* that lifted a proposition for *A* to *Free A*.

Inductive *ForFree A* (P : A → Prop) : Free $C_F$ A → Prop :=
| *For_pure* : ∀ x : A, P x → ForFree P (pure x)
| *For_impure* : ∀ (s : Shape) (pf : Pos s → Free $C_F$ A),
    (∀ p : Pos s, ForFree P (pf p)) → ForFree P (impure (ext s pf))

For a given proposition P : A → Prop Mona distinguished two cases. In case of *pure x* the proposition was valid if *P x* held. For *impure (ext s pf)* the proposition was valid if *P* held for all elements of the container *ext s pf*. To state that *P* held for all elements of the container *ext s pf* Mona used the same idea as for the definition of *Free_Ind*.

With the definition of *ForFree* at hand, she defined the induction principle *List_Ind* that used a hypothesis about the remaining list in the *cons* case.

*List_Ind* : . . .
  . . .
  → (∀ (fx : Free $C_F$ A) (fxs : Free (List $C_F$ A)), ForFree P fxs → P (cons fx fxs))
  . . .

One detail Mona did not like about these definitions was that the induction hypothesis generated by *List_Ind* seemed so complicated to use. The hypothesis was of the form *H* : *ForFree P* (*pure xs*), where *P* was the statement Mona needed to conclude the proof. She could derive from the definition of *ForFree* that *P* held for *xs*, but she had to apply a lot of tactics to get access to that hypothesis. In order to make the usage of *List_Ind* more convenient, Mona wrote a shortcut that brought the induction hypothesis into the required form and could also handle such a hypothesis for the *impure* case. The shortcut *simplify H* as *IH* simplified a hypothesis *H* generated by *List_Ind* and introduced the required hypothesis under the name *IH*.

## About the authors

**Sandra Dylus** is a PhD student in the group for programming languages and compiler construction at the University of Kiel. Contact her at sad@informatik.uni-kiel.de.

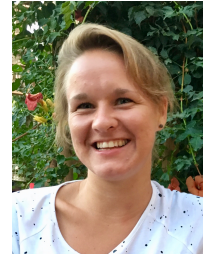

**Jan Christiansen** is a Professor for programming languages and the theory of programming at the University of Applied Sciences Flensburg. Contact him at jan.christiansen@hs-flensburg.de

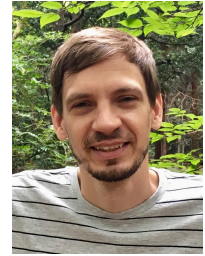

**Finn Teegen** is also a PhD student in the group for programming languages and compiler construction at the University of Kiel. Contact him at fte@informatik.uni-kiel.de

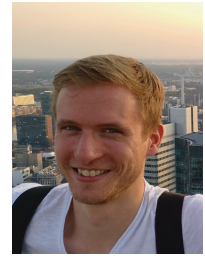